\documentclass[prd,aps,twocolumn]{revtex4}

\usepackage{graphicx}
\usepackage{dcolumn}
\usepackage{bm}

\newcommand{\plb}[2]{{\em Phys. Lett.}              {\bf #1B}, #2 }
\newcommand{\npb}[2]{{\em Nucl. Phys.}              {\bf B#1}, #2 }

\newcommand{\pr }[2]{{\em Phys. Rep.}               {\bf  #1}, #2 }
\newcommand{\prt}[2]{{\em Phys. Rev.}               {\bf D#1}, #2 }

\newcommand{\mpl}[2]{{\em Mod. Phys. Lett.}         {\bf A#1}, #2 }

\newcommand{\be}{\begin{equation}}
\newcommand{\ee}{\end{equation}}
\newcommand{\ba}{\begin{eqnarray}}
\newcommand{\ea}{\end{eqnarray}}
\newcommand{\bt}{\begin{table}}
\newcommand{\et}{\end{table}}
\newcommand{\brt}{\begin{ruledtabular}}
\newcommand{\ert}{\end{ruledtabular}}
\newcommand{\btu}{\begin{tabular}}
\newcommand{\etu}{\end{tabular}}

\def\tr{{\rm Tr}}

\def\th{\theta}
\def\ka{\kappa}
\def\la{\lambda}
\def\cmn{\chi_{mn}}
\def\cpq{\chi_{pq}}
\def\lt{\left}
\def\rt{\right}
\def\ov{\over}
\def\nonu{\nonumber}
\def\QSmn{Q^{2S}_{mn}}
\def\QSpq{Q^{2S}_{pq}}
\def\QSmp{Q^{2S}_{mp}}
\def\QSnq{Q^{2S}_{nq}}
\def\QSnp{Q^{2S}_{np}}
\def\QFmn{Q^{2F}_{mn}}
\def\QFpq{Q^{2F}_{pq}}
\def\QFmp{Q^{2F}_{mp}}
\def\QFnq{Q^{2F}_{nq}}
\def\QFnp{Q^{2F}_{np}}

\def\sump{\sum_{perms}}

\begin{document}

\preprint{
\noindent
\hfill
\begin{minipage}[t]{3in}
\begin{flushright}
\vspace*{2cm}
\end{flushright}
\end{minipage}
}


\title{Renormalization Group Equations in Gauge Theories with Multiple $U(1)$ Groups}
\author{Mingxing Luo and Yong Xiao}
\affiliation{Zhejiang Institute of Modern Physics, Department of Physics,
Zhejiang University, Hangzhou, Zhejiang 310027, P R China}

\date{\today}

\begin{abstract}
Two-loop renormalization group equations in gauge theories 
with multiple $U(1)$ groups are presented.
Instead of normalizing the abelian gauge fields in canonical forms, 
we retain kinetic-mixing terms and treat the mixing coefficients as free parameters.
The $\beta$- and $\gamma$-functions are then obtained in a straightforward manner.
\end{abstract}


\maketitle

\section{Introduction}
Renormalization group equations (RGEs) relate parameters at different scales.
It plays an important role in the search of physics beyond the standard model (SM)
and provides a unique window for physics at extremely high energy.
Comprehensive study of two-loop RGEs in a general gauge theory
had been performed long ago \cite{MV}.
The calculation was first performed for a simple Lie group.
By using suitable substitution rules, these results can be extended to cases
of semi-simple Lie groups, as well as cases with only one $U(1)$ sector.
Recently, these calculations were updated by carefully treating the
transformation properties of the fermion fields \cite{xiao} 
and applied to the special case of the SM\cite{luo}.
Extension to cases with multiple $U(1)$ groups is more involved, due to 
the possible presence of kinetic mixing terms between different $U(1)$ gauge bosons.
In this letter, we extend the calculation to these cases.

Gauge theories with multiple $U(1)$ groups can arise in many context.
In most grand unified theories, one in general has more than one $U(1)$ sector.
In string theories, one obtains extra $U(1)$ sectors copiously.
And there are no general rules to make these extra $U(1)$ gauge bosons heavy.
Were they discovered, extra $U(1)$ bosons would provide 
important signatures of the underlying theory.
In some model buildings, extra gauge $U(1)$ bosons
were sought to solve the $\mu$ problem in supersymmetric theories \cite{mu}. 
It is arguably that extra $U(1)$ bosons are the most likely new physics next to 
supersymmetry to be discovered.
Over the years, models with extra $U(1)$'s have been extensively studied \cite{SMU1}.

Unique to multiple $U(1)$ gauge theories, there could be kinetic mixing terms in the form
\be
- {1\ov4} \sum_{m\neq n} \xi_{mn} F^{m\mu\nu}F^n_{\mu\nu} 
\ee
which are both Lorentz and gauge invariant.
One needs extra symmetries to keep them out, while in non-abelian gauge theories,
similar terms are forbidden automatically by gauge invariance.
Even if one diagonalizes the the abelian gauge fields 
such that $\xi_{mn}(m\neq n)$ vanish at one scale,
the mixing terms will in general come back at another scale due to RGE runnings \cite{dienes}.
Especially, 
these mixings affect the various $\beta$- and $\gamma$-functions,
which should be included in consistent analysis.
Conventionally, one normalizes the abelian gauge fields into canonical forms, 
such that the kinetic mixing terms vanish at the tree level.
For example, the $\beta$-functions of gauge couplings have been obtained up to two-loops
via this procedure in  \cite{aguila}.
However, mixings usually come back in loop-levels 
and the whole procedure becomes rather complicated. 
In this letter, we will provide an alternative approach. 
We retain the kinetic-mixing terms and treat $\xi_{mn}$ as free parameters.
It proves to be extremely convenient in practice.
The $\beta$- and $\gamma$-functions of the theory, 
including the $\beta$-functions of the $\xi_{mn}$'s,
can be obtained in a straightforward manner by extending existing calculations.

The paper is organized as follows,
section 2 presents our formalism and the $\gamma$-functions of the scalar and fermion fields.
Section 3 presents the $\beta$-functions of the gauge couplings and those of the $\xi_{mn}$'s.
Section 4 presents the $\beta$-functions of the Yukawa couplings 
and section 5 those of quartic scalar couplings.
We conclude in section 6.

\section{formalism}
We start with a general renormalizable field theory with a group 
$G \times \prod_m U_m(1)$, where $G$ is a compact simple Lie group.
The theory contains real scalar fields $\phi_a$ and two-component fermion fields $\psi_j$.
The Lagrangian can be written as
\be
{\cal L} = {\cal L}_0  + ({\rm gauge\ fixing + ghost\ terms})
\ee
where
\ba
{\cal L}_0 &= &  - {1\ov4} \sum_{mn} \xi_{mn} F^{m\mu\nu} F^n_{\mu\nu} 
           -  {1\ov4} F^{A \mu\nu} F_{\mu\nu}^A     \nonu \\ 
           &+& {1\ov2}D^\mu \phi_a  D_\mu \phi_a
                 + i\psi_j^+ \sigma^\mu D_\mu \psi_j  \nonu \\
            &-& {1\ov2} \lt( Y_{jk}^a \psi_j \zeta \psi_k \phi_a + h.c. \rt)
                - {1\ov4!} \la_{abcd} \phi_a \phi_b \phi_c \phi_d,
\ea
where the diagonal terms with $\xi_{mm}=1$ correspond to the usual kinetic terms for
abelian gauge bosons, 
while the off-diagonal $\xi_{mn}(m\neq n)$ represent kinetic mixings.
The gauge field strengths are
\ba
F^m_{\mu\nu}= \partial_\mu V^m_\nu -  \partial_\nu V^m_\mu, \nonu \\
F^A_{\mu\nu}= \partial_\mu V^A_\nu -  \partial_\nu V^A_\mu + g f^{ABC} V^B_\mu V^C_\nu,  \nonu
\ea
and covariant derivatives of matter fields are
\ba
D_\mu \phi_a &=& \partial_\mu \phi_a - ig\th_{ab}^A V_\mu^A \phi_b
              - i \sum_m g_m (Q_m^S)_{ab} V^m_{\mu} \phi_b, \nonu \\
D_\mu \psi_j &=& \partial_\mu \psi_j - igt_{jk}^A V_\mu^A \psi_k 
              - i \sum_m g_m (Q_m^F)_{jk} V^m_{\mu} \psi_k, \nonu
\ea
where $\th^A$ and $t^A$ are the representation matrices of $G$ on the scalar and fermion fields,
respectively, and $Q_m^S$ and $Q_m^F$ are their $U_m(1)$ charges.
Since the scalar fields are real, $\th^A$ and $Q_m^S$ are pure imaginary and antisymmetric.
$Q_m^F$ are real and diagonal.
The constraint on Yukawa coupling matrices imposed by $U(1)$ gauge invariance is
\be
Y^b (Q_m^S)_{ba} + Y^a Q_m^F + Q^{F}_m Y^a =0,
\ee
which can be used to simplify the gauge structures.

Conventionally, one normalizes the $U(1)$ gauge fields to assume canonical forms 
before renormalization.
It is easy to show that this normalization does not affect 
the $\beta$-functions of $g$, $Y^a$ and $\la_{abcd}$
and the $\gamma$-functions of $V^A$, $\phi_a$ and $\psi_j$ \cite{leeyang,luo2}.
On the other hand, the $\beta$-functions of $g_m$ in the conventional scheme are
related to the ones in our scheme by a $\xi$-dependent transformation.
If one normalizes the $U(1)$ gauge fields consistently, 
both schemes give the same physics prediction \cite{luo2}.

In the Landau gauge, the field propagators of the simple gauge group $G$ are 
\be
D_{\mu\nu}^{AB}(k) = \delta^{AB} \left( - g_{\mu\nu} + {k_\mu k_\nu \over k^2} \right)
{i \over k^2},
\ee
and the propagators of abelian gauge fields are
\ba
D_{\mu\nu}^{mn}(k) 
&=& \xi^{-1}_{mn} \left( - g_{\mu\nu} + {k_\mu k_\nu \over k^2} \right) {i \over k^2},
\label{propagator}
\ea
where $\xi^{-1}_{mn}$ is the inverse of $\xi_{mn}$. 
Note that, there are propagations from a $U_m(1)$ gauge field to a $U_n(1)$ gauge
field due to kinetic mixings, with magnitude proportional to $\xi^{-1}_{mn}$.
For late convenience, we define
\ba
C_2^{ab} (S) = \th^A_{ac} \th^A_{cb}, \ S_2(S) \delta^{AB} = \tr [\th^A \th^B], \\
C_2^{ab} (F) = t^A_{ac} t^A_{cb}, \ S_2(F) \delta^{AB} = \tr [t^A t^B], \\
C_2(G) \delta^{AB} = f^{ACD} f^{BCD}, \\
Q_{mn}^{2S} = Q_m^S Q_n^S, \ Q_{mnpq}^{4S} = Q_m^S Q_n^S Q_p^S Q_q^S, \\
Q_{mn}^{2F} = Q_m^F Q_n^F, \ Q_{mnpq}^{4F} = Q_m^F Q_n^F Q_p^F Q_q^F, \\
\chi_{mn} = g_m g_n \xi^{-1}_{mn},
\label{chimn}
\ea
where in Eq.~(\ref{chimn}), there is no summation over $m$ and $n$.

Except for the $\xi_{mn}(m\neq n)$, 
we separate the $\beta$-function of each coupling constant $x$ into two parts,
\be
 \beta_x = \beta_x(G)+\beta_x(U),
\ee
where $\beta_x(G)$ is the result when one turns off all $U(1)$ interactions and
$\beta_x(U)$ is the additional contribution when the $U(1)$ interactions are turned on.
To two-loops, we have
\ba
\beta_x &=& {1\over16\pi^2} \lt[\beta_x^{(1)}(G)+\beta_x^{(1)}(U)\rt] \nonu \\
&+& {1\over(16\pi^2)^2} \lt[\beta_x^{(2)}(G)+\beta_x^{(2)}(U)\rt].
\ea
Similarly, we separate the $\gamma$-function of field $i$
\ba
\gamma_i &=& {1 \over (16 \pi^2)} \lt[\gamma_i^{(1)}(G)+\gamma_i^{(1)}(U)\rt] \nonu \\
&+& {1 \over (16 \pi^2)^2} \lt[\gamma_{i}^{(2)}(G)+\gamma_i^{(2)}(U) \rt].
\ea
To two-loops, $\beta(G)$ and $\gamma(G)$ have been given in \cite{MV,xiao}.
The additional contributions from $U(1)$ groups, 
$\beta_x(U)$ and $\gamma_i(U)$ are to be calculated in this letter.
These can be obtained from existing calculations
by closely inspecting the Feynman diagrams in \cite{MV},
and making proper substitution of gauge couplings and abelian gauge propagators.
The process is straightforward but tedious. 
We content with by simply reporting the end results here.

\subsection{Scalar wave function renormalization}
To one loop, the additional contributions to the $\gamma$-functions of
scalar fields induced by $U(1)$ groups are 
\be
\gamma_{ab}^{S(1)}(U) = - 3\cmn (\QSmn)_{ab},
\ee
here and hereafter repeated indices such as $m$ and $n$ are
implicitly summed over all abelian gauge groups, unless specified otherwise.
To two loops, the additional contributions are
\ba
\gamma_{ab}^{S(2)}(U) &=& \cmn \left\{
5\ka \tr \lt[ \QFmn (Y^{+a} Y^b + Y^{+b} Y^a) \rt] \right. \nonu \\
&+& {1\ov12} \cpq (\QSmp)_{ab} [11 \tr (\QSnq) 
+ 40 \ka \tr (\QFnq) ] \nonu \\
&+& {3\ov2} \cpq (Q^{4S}_{mnpq})_{ab}
+ \left. 3 g^2 \lt[C_2(S) \QSmn \rt]_{ab} \right\},
\ea
where $\ka={1\ov2}(1)$ for two(four)-component fermions.

\subsection{Fermion wave function renormalization}
To one loop, the additional contributions to the $\gamma$-functions of
fermion fields induced by abelian groups are zero.
To two loops,
\ba
\gamma_F^{(2)}(U) &=&
 \cmn \left\{- {1\ov4} Y^a \QFmn Y^{+a} \nonu \right. \\
&-& {7\ov4}  \QFmn Y^a Y^{+a}
+ {9\ov2} (\QSmn)_{ab} Y^a Y^{+b} \nonu \\
&-& {1\ov4}  \cpq (\QFmp) [ \tr (\QSnq) 
+ 8 \ka  \tr (\QFnq)] \nonu \\
&-& {3\ov2}  \cpq Q^{4F}_{mnpq} 
- \Big. 3 g^2  C_2(F) \QFmn \Big\}.
\ea

\section{$\beta$-functions of gauge couplings and kinetic mixings}
\subsection{Nonabelian gauge coupling}
At one loop level, the additional contributions to the $\beta$-function of
gauge coupling $g$ of group $G$ are zero. To two loop, one has
\be
\beta_g^{(2)}(U) = 2 g^3 \cmn T_{mn}^G (S) + 4 \ka g^3 \cmn T_{mn}^G (F)
\label{nonabelian}
\ee
where $T_{mn}^G(S)$ and $T_{mn}^G(F)$ are defined as
\ba
T_{mn}^G(S) \delta^{AB} &=& \tr (\th^A \th^B \QSmn) \nonu \\
T_{mn}^G(F) \delta^{AB} &=& \tr (t^A t^B \QFmn) \nonu
\ea

\subsection{Abelian gauge couplings}
The $\beta$-function of the gauge coupling $g_p$ of group $U_p(1)$ is 
related to the anomalous dimension $\gamma_p$ of the corresponding gauge field $V_\mu^p$[Figure 1(a)].
The Ward identity ensures that $\beta_p =  g_p \gamma_p$.
To two loops,
\ba
\beta_{p} =
&-& {g_p^3\ov(4\pi)^2} \lt\{ -{4\ov3}\ka S_2^p(F) - {1\ov6}S_2^p(S) +
                         {2\ka\ov(4\pi)^2} Y_4^p(F) \rt\} \nonu \\
&+& {g_p^3\ov(4\pi)^4} \lt\{ 2 \cmn \left[ \tr(Q^{4S}_{ppmn}) 
    + 2  \ka \tr(Q^{4F}_{ppmn}) \rt] \rt. \nonu \\
&+& \left. 2 g^2 C_2(S) S_2^p (S) + 4 \ka  g^2 C_2(F) S_2^p (F) \rt\}
\ea
where
\be
Y_4^p(F) = \tr \lt[ C_2^p(F) Y^a Y^{+a} \rt],
\ee
and there is no summation over $p$.
$C_2(R)$ is the Casimir operator of $G$ on the representations $R$.
$C_2^p(R)$ and $S_2^p(R)$ are the Casimir operator
and the Dykin index of $U_p(1)$ acting on fields $R$, respectively.

\subsection{Kinetic mixings}
The $\beta$-functions of the kinetic mixing coefficients $\xi_{pq}$ are given by
\be
\beta_{pq} = (\gamma_p + \gamma_q) \xi_{pq} + \gamma^\xi_{pq}
\ee
where repeated indices don't imply summation and
$\gamma^\xi_{pq}$ is the anomalous dimension of the operator $F^{p\mu \nu} F^q_{ \mu \nu}$[Figure 1(b)],
\ba
\gamma^\xi_{pq} &=&
  -  {g_p g_q\ov(4\pi)^2}
\lt[ {1\ov3} \tr(\QSpq) + {8\ov3} \ka \tr(\QFpq) \rt] \nonu \\
&-& {g_p g_q\ov(4\pi)^4}
\Big\{ -4 \ka \tr(Y^a Y^{+a} \QFpq) \Big. \nonu \\
&+& 4 g^2 \left( \tr \lt[\QSpq C_2(S)\rt]
    + 2 \ka \tr \lt[ \QFpq C_2(F) \rt] \right) \nonu \\
&+& 4 \cmn \lt( \tr(Q^{4S}_{pqmn})
   + \Big. 2 \ka \tr(Q^{4F}_{pqmn}) \rt) \Big\}
\ea
where the indices are summed over $m(n)$ but not over $p(q)$.
\begin{figure}
\includegraphics[width=3.5in]{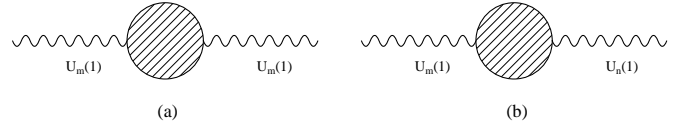}
\caption{\label{fig:vector}(a)Wave function renormalization of abelian gauge fields;
(b)renormalization of abelian gauge kinetic mixing.}
\end{figure}

\section{Yukawa coupling}
The $\beta$-functions of the Yukawa couplings can be expressed as
\be
\beta^a = \gamma^a + \gamma^{+F} Y^a + Y^a \gamma^F + \gamma_{ab}^S Y^b
\ee
where $\gamma^a$ are the anomalous dimensions of the operators $\phi_a \psi_j \zeta \psi_k$,
$ \gamma^F$ and $\gamma_{ab}^S$ are the anomalous dimensions of the corresponding fermions
and bosons, respectively.
To one loop level, the additional contributions to the $\beta$-function of Yukawa
couplings are
\be
\beta_a^{(1)}(U) = - 3 \cmn (\QFmn Y^a + Y^a \QFmn)
\ee
To two loop level, one has
\ba
\beta_a^{(2)}(U) 
&=& \cmn \Big\{ 3 \{ \QFmn, Y^b Y^{+a} Y^b \} \Big. 
+ 5  Y^b \{ \QFmn, Y^{+a} \} Y^b \nonu \\
&-& {7\ov4}  \lt[ \QFmn Y^b Y^{+b} Y^a + Y^a Y^{+b} Y^b \QFmn \rt] \nonu \\
&-& {1\ov4}  (Y^b \QFmn Y^{+b} Y^a + Y^a Y^{+b} \QFmn Y^b) \nonu \\
&+& 6  (Q_m^{F} Y^a Y^{+b} Q_n^{F} Y^b + Y^b Q_m^F Y^{+b} Y^a Q_n^F) \nonu \\
&+& 5\ka  Y^b \tr \lt[ \QFmn (Y^{+a} Y^b + Y^{+b} Y^a) \rt] \nonu \\
&+& 6  \lt[ (\QSmn)_{bc} Y^b Y^{+a} Y^c -2 (\QSmn)_{ac} Y^b Y^{+c} Y^b \rt] \nonu \\
&+& {9\ov2}  (\QSmn)_{bc} (Y^b Y^{+c} Y^a + Y^a Y^{+c} Y^b) \nonu \\
&-& 3 g^2  \{ \QFmn C_2(F), Y^a\} - {3\ov2}  \cpq \{ Q_{mnpq}^{4F}, Y^a \} \nonu \\
&+& 6 g^2  C_2^{ab}(S) \{ \QFmn, Y^b\} + 6 g^2  (\QSmn)_{ab} \lt\{ C_2(F), Y^b \rt\} \nonu \\
&+& 6  \cpq (\QSmn)_{ab} \{ \QFpq, Y^b \} \nonu \\
&+& {1\ov12}  \cpq \{ \QFmp, Y^a \} [ 11 \tr(\QSnq) + 40 \tr(\QFnq) ]\nonu \\
&-& 21 \left[ g^2  (\QSmn C_2(S))_{ab}  
    + {1\ov2}  \cpq (Q^{4S}_{mnpq})_{ab} \right] Y^b \nonu \\
&-& {1\ov4} \cpq (\QSmp)_{ab} \left[ 8 \ka \tr(\QFnq) 
   + \Big. \tr(\QSnq) \right] Y^b \Big\}
\ea

\section{Scalar quartic coupling}
The $\beta$-functions of the scalar quartic couplings can be expressed as
\be
\beta_{abcd} = \gamma_{abcd} + \sum_i \gamma^S(i) \la_{abcd}
\ee
where $\gamma_{abcd}$ are the anomalous dimensions of the operators
$\phi_a \phi_b \phi_c \phi_d$, $ \gamma^S(i)$ is the anomalous dimension of
the scalar field $i$.
To one loop level, one has
\ba
\beta_{abcd}^{(1)}(U) =
&-&3 \cmn \Lambda^{Q,mn}_{abcd} + 6 g^2 \cmn A^{mn}_{abcd} \nonu \\
&+& 3 \cmn \cpq A^{mnpq}_{abcd}
\ea
and to two loop level, one has
\ba
\beta_{abcd}^{(2)}(U)
&=& \cmn \lt[ 2 \bar{\Lambda}^{mn}_{abcd} - 6 \Lambda^{mn}_{abcd} \rt. \nonu \\
&+& 4 \ka ( H^{Q,mn}_{abcd} - H^{F,mn}_{abcd} )
     \lt. + 5 \ka \sum_i Y^{Q2F}(i) \la_{abcd} \rt] \nonu \\
&+& \cmn \cpq \lt\{ \lt[ {10\ov3} \ka \tr(\QFnq) \rt. \rt. \nonu \\
&+& \lt. {11\ov12} \tr(\QSnq) \rt] \Lambda^{Q,mp}_{abcd}  \nonu \\
&+& {3\ov2} \Lambda^{2Q,mnpq}_{abcd} + {5\ov2} A^{\la,mnpq}_{abcd}
     + {1\ov2} \bar{A}^{\la,mnpq}_{abcd}  \nonu \\
&-& \lt. 4\ka(B^{Y,mnpq}_{abcd}-10\bar{B}^{Y,mnpq}_{abcd}) \rt\} \nonu \\
&+& g^2 \cmn \lt[ 3 \Lambda^{SQ,mn}_{abcd} + 5 A^{\la,mn}_{abcd} \rt. \nonu \\
&+&  \lt. \bar{A}^{\la,mn}_{abcd}
     - 8\ka (B^{Y,mn}_{abcd} - 10 \bar{B}^{Y,mn}_{abcd}) \rt] \nonu \\
&-& \cmn \cpq \chi_{kl} \lt\{ {15\ov2} A^{mnpqkl}_{abcd} \rt. \nonu \\
&+& \lt[{7\ov3} \tr(\QSnp) + {32\ov3} \ka \tr(\QFnp) \rt]
    \lt. A^{mqkl}_{abcd} \rt\} \nonu \\
&-& g^2 \cmn \cpq \lt\{ \rt.
    15 A^{Q,mnpq}_{abcd} + {15\ov2} A^{S,mnpq}_{abcd} \nonu \\
&+&  \lt[{7\ov3} \tr(\QSnp) + {32\ov3} \ka \tr(\QFnp) \rt]
     \lt. A^{mq}_{abcd} \rt\} \nonu \\
&-& g^4 \cmn \lt\{ {15\ov2}A^{Q,mn}_{abcd} + 15 A^{S,mn}_{abcd} \rt. \nonu \\
&+& \lt. \lt[ {7\ov3}S_2(S)
    + {32\ov3}\ka S_2(F) - {161\ov6} C_2(G)\rt] A^{mn}_{abcd} \rt\} \nonu \\
\ea
The group factors are defined as
\ba
\Lambda^{Q,mn}_{abcd}
&=& \sum_i C_2^{Q,mn}(i) \la_{abcd} \nonu  \\
\bar{\Lambda}^{mn}_{abcd}
&=& {1\ov8} \sump (\QSmn)_{fg} \la_{abef} \la_{cdeg} \nonu  \\
\Lambda^{mn}_{abcd}
&=& {1\ov8} \sump \la_{abef} \la_{cdgh} (Q_m^S)_{eg} (Q_n^S)_{fh} \nonu  \\
H^{Q,mn}_{abcd}
&=& \sum_i C_2^{Q,mn}(i) H_{abcd} \nonu  \\
H^{F,mn}_{abcd}
&=& \sump \tr \lt[ \{ \QFmn, Y^a \} Y^{+b} Y^c Y^{+d} \rt]  \nonu \\
\Lambda^{2Q,mnpq}_{abcd}
&=& \sum_i C_2^{Q,mn}(i) C_2^{Q,pq}(i) \la_{abcd} \nonu  \\
\Lambda^{SQ,mn}_{abcd}
&=& \sum_i C_2(i) C_2^{Q,mn}(i) \la_{abcd} \nonu  \\
Y^{Q2F} \delta^{ab}
&=& \tr \lt[ \QFmn(Y^{+a} Y^b + Y^{+b} Y^a)\rt] \nonu  \\
A^{\la,mnpq}_{abcd}
&=& {1\ov4} \sump \la_{abef} \{Q_m^S,Q_p^S\}_{ef} \{Q_n^S,Q_q^S\}_{cd} \nonu  \\
A^{\la,mn}_{abcd}
&=& {1\ov4} \sump \la_{abef} \{Q_m^S,\th^A\}_{ef} \{Q_n^S,\th^A\}_{cd} \nonu  \\
\bar{A}^{\la,mnpq}_{abcd}
&=& {1\ov4} \sump \la_{abef} \{Q_m^S,Q_p^S\}_{ce} \{Q_n^S,Q_q^S\}_{df} \nonu  \\
\bar{A}^{\la,mn}_{abcd}
&=& {1\ov4} \sump \la_{abef} \{Q_m^S,\th^A\}_{ce} \{Q_n^S,\th^A\}_{df} \nonu \nonu  \\
B^{Y,mnpq}_{abcd}
&=& {1\ov4} \sump \{Q_m^S,Q_p^S\}_{ab} \tr \lt( Q_n^{F} Q_q^{F} Y^c Y^{+d} \rt. \nonu \\
&+& \lt. Y^c Q_n^F Q_q^F Y^{+d} \rt) \nonu  \\
B^{Y,mn}_{abcd}
&=& {1\ov4} \sump \{Q_m^S,\th^A\}_{ab} \tr \lt( Q_n^{F*} t^{A*} Y^c Y^{+d} \rt. \nonu \\
&+& \lt. Y^c Q_n^F t^A Y^{+d} \rt) \nonu  \\
\bar{B}^{Y,mnpq}_{abcd}
&=& {1\ov4} \sump \{Q_m^S,Q_p^S\}_{ab} \tr(Q_n^{F} Y^c Q_q^F Y^{+d}) \nonu  \\
\bar{B}^{Y,mn}_{abcd}
&=& {1\ov4} \sump \{Q_m^S,\th^A\}_{ab} \tr(Q_n^{F} Y^c t^A Y^{+d}) \nonu  \\
A^{mn}_{abcd}
&=& {1\ov8} \sump \{\th^A,Q_m^S\}_{ab} \{\th^A,Q_n^S\}_{cd} \nonu  \\
A^{mnpq}_{abcd}
&=& {1\ov8} \sump \{Q_m^S,Q_p^S\}_{ab} \{Q_n^S,Q_q^S\}_{cd} \nonu  \\
A^{mnpqkl}_{abcd}
&=& \sum_i C_2^{Q,mn}(i) A^{pqkl}_{abcd} \nonu  \\
A^{Q,mn}_{abcd}
&=& \sum_i C_2^{Q,mn}(i) A_{abcd} \nonu  \\
A^{Q,mnpq}_{abcd}
&=& \sum_i C_2^{Q,mn}(i) A^{pq}_{abcd} \nonu  \\
A^{S,mn}_{abcd}
&=& \sum_i C_2(i) A^{mn}_{abcd} \nonu  \\
A^{S,mnpq}_{abcd}
&=& \sum_i C_2(i) A^{mnpq}_{abcd} \nonu 
\ea
where $C_2^{Q,mn}(k)$ are the eigenvalues of $Q_m^S Q_n^S$, and
\ba
A_{abcd} &=&
{1\ov8} \sump \{\th^A,\th^B\}_{ab} \{\th^A,\th^B\}_{cd} \nonu  \\
H_{abcd} &=& \sump {1\ov4} \tr(Y^a Y^{+b} Y^c Y^{+d}) \nonu 
\ea

\section{Conclusion}
We have now presented the two-loop RGEs
for all dimensionless parameters in a general gauge theory 
with multiple abelian gauge groups.
We have retained all kinetic mixing terms 
and did not canonically normalize the abelian gauge fields.
This approach proved to be extremely convenient.
Were all abelian gauge interactions turned off,
the two-loop RGEs were given in \cite{MV,xiao}.
When the abelian gauge interactions are turned on,
the additional contributions have been obtained from existing calculations
by a close inspection of the relevant Feynman diagrams and
by using suitable substitution rules. 
Were $G$ not simple, but rather a direct product of simple groups $G_k$ with
gauge coupling $g_k$,
the RGEs can again be obtained by using substitution rules as given in \cite{MV,xiao}.
For part of $g^4$-related terms, the following extra substitution rules are needed,
\ba
g^4 A^{Q,mn}_{abcd} \rightarrow 
{1\ov8} g_k^2 g_l^2 \sum_i C_2^{Q,mn}(i) 
\sump \{\th^A_k,\th^B_l\}_{ab} \{\th^A_k,\th^B_l\}_{cd} \nonu \\
g^4 A^{S,mn}_{abcd} \rightarrow 
{1\ov8} g_k^2 g_l^2 \sum_i C_2^k(i) 
\sump \{\th^A_l,Q_m^S\}_{ab} \{\th^A_l,Q_n^S\}_{cd} \nonu \\
g^4 S_2(R) A^{mn}_{abcd} \rightarrow 
{1\ov8} g_k^4 S_2^k(R) 
\sump \{\th^A_k,Q_m^S\}_{ab} \{\th^A_k,Q_n^S\}_{cd} \nonu \\
g^4 C_2(G) A^{mn}_{abcd} \rightarrow 
{1\ov8} g_k^4 C_2^k(G) 
\sump \{\th^A_k,Q_m^S\}_{ab} \{\th^A_k,Q_n^S\}_{cd} \nonu \\
\ea
By introducing a dummy field, $\beta$-functions of the dimensional parameters
can also be deduced from those of $Y^a$ and $\la_{abcd}$ \cite{xiao,luo2}.

\begin{acknowledgments}
The work was supported by a Fund for Trans-Century Talents,
CNSF-90103009  and CNSF-10047005.
\end{acknowledgments}

\end{document}